\input{aipcheck}

\documentclass[
    ,final            
  ]
  {aipproc}

\layoutstyle{6x9}

\begin{document}

\title{Nonleptonic $B_s$ to charmonium decays and their role in the determination of the $\beta_s$}


\author{Wei Wang}{
  address={Istituto Nazionale di Fisica Nucleare, Sezione di Bari, Via Orabona 4, I-70126 - Bari, Italy.}
,altaddress={Istituto Nazionale di Fisica Nucleare, Sezione di Bari, Via Orabona 4, I-70126 - Bari, Italy.} 
}

\begin{abstract}
This talk consists of two parts.  We first present a light-cone QCD
sum rule computation of the $B_s\to f_0(980)$ form factors which are
necessary inputs in semileptonic and nonleptonic $B_s$ decays into
$f_0(980)$. Then we analyze nonleptonic $B_s$ decays into a
charmonium state and a light meson, which are potentially useful to
access the $B_s$-${\bar B}_s$ mixing phase $\beta_s$.  We explore
the experimental feasibility of measuring these  various
channels, paying attention to different determinations of $\beta_s$
in view of the hints of new physics recently emerged in the $B_s$
sector.
\end{abstract}

\maketitle

\section{Introduction}

The detailed study of CP violation is a powerful and rigorous tool in the
discrimination between the Standard Model (SM) and alternative
scenarios. For instance the analysis of the $B_s$ unitarity
triangle of the
Cabibbo-Kobayashi-Maskawa (CKM) matrix elements:
$V_{us}V_{ub}^*+V_{cs}V_{cb}^*+V_{ts}V_{tb}^*=0$ provides an
important test of the SM description of CP violation. One of its
angles, defined as $\beta_s=Arg\big[-{V_{ts}V_{tb}^* \over V_{cs}
V_{cb}^*}\big]$, is half of the phase in the $B_s$-${\bar B}_s$
mixing, and is expected to be tiny in the SM: $\beta_s \simeq 0.017$
rad.  The current measurements, by the CDF and D{{\O}}
collaborations at Tevatron based on the angular analysis of the
time-dependent differential decay width in the process $B_s \to
J/\psi \phi$~\cite{:2008fj}, indicate larger values and the averaged
results are consistent with SM only at 2.2 $\sigma$ level:
$\phi_s^{J/\psi \phi}=-2 \beta_s=-0.77 \pm^{0.29}_{0.37}$ or
$\phi_s^{J/\psi \phi}=-2 \beta_s=-2.36
\pm^{0.37}_{0.29}$~\cite{Barberio:2008fa}. Although the recent
result by the CDF: $\beta_s \in [0.0,0.5] \,{\rm U} \,[1.1,1.5]$ (at
68$\%$ confidence level) \cite{talkcdf} has a smaller deviation from
the SM, the uncertainties are still large and the precise
measurement of $\beta_s$ is a priority for the forthcoming
experiments. Towards this direction the nonleptonic $B_s$ decays are
certainly of prime importance.

In this work we first  compute the $B_s \to f_0(980)$ form factors
using the light-cone QCD sum rule
(LCSR)\cite{Colangelo:2010bg}. These results will be useful in the
analysis of semileptonic and nonleptonic $B_s\to f_0$ decays. Subsequently we investigate the $B_s$ decay modes
induced by the transition $b \to c {\bar c}s$, namely $B_s \to
M_{c{\bar c}} + L$, where $M_{c{\bar c}}$ is an s-wave or p-wave
charmonium state and $L$ is a light scalar, pseudoscalar or vector
meson, $f_0(980)$, $\eta$, $\eta^\prime$,
$\phi$~\cite{Colangelo:2010zz}. In particular, we exploit the
generalized factorization approach to calculate their branching
fractions in the SM in order to understand which of these modes
are better suitable to determine $\beta_s$.

\section{ $B_s\to f_0$ form factors in LCSR \label{sec:formfactors-LCSR}}

Hereafter we will use $f_0$ to denote $f_0(980)$ meson for simplicity. The
parametrization of matrix elements involved in $B_s\to f_0$
transitions is expressed in terms of the form factors
\begin{eqnarray}
 &&\langle f_0(p_{f_0})|J^5_\mu |\overline {B}_s(p_{B_s})\rangle= -i\big\{F_1(q^2)\big[P_\mu
 -\frac{m_{B_s}^2-m_{f_0}^2}{q^2}q_\mu\big]+ F_0(q^2)\frac{m_{B_s}^2-m_{f_0}^2}{q^2}q_\mu\big\}, \label{F1-F0}
 \nonumber\\
  &&\langle{f_0}(p_{f_0})|J_\mu^{5T}|\overline {B}_s(p_{B_s})\rangle
 =-\frac{F_T(q^2)}{m_{B_s}+m_{f_0}}   \Big[q^2P_\mu
 -(m_{B_s}^2-m_{f_0}^2)q_\mu\Big], \label{FT}
\end{eqnarray}
where $P=p_{B_s}+p_{f_0}$, $q=p_{B_s}-p_{f_0}$,  and $J_\mu^5=\bar
s\gamma_\mu\gamma_5b$,  $J_\mu^{5T}=\bar
s\sigma_{\mu\nu}\gamma_5 q^\nu b$ . To compute such form factors in
the LCSR~\cite{Colangelo:2000dp} we consider the correlation
function:
\begin{eqnarray}
  \Pi(p_{f_0},q)= i \int d^4x \, e^{iq\cdot x} \langle {f_0}(p_{f_0})|{\rm
 T}\left\{j_{\Gamma_1}(x),j_{\Gamma_2}(0)\right\}|0\rangle \hspace{-0.1cm}
 \label{corr}
\end{eqnarray}
with $j_{\Gamma_1}$ being one of the currents in the definition of
the $B_s\to f_0$ form factors: $j_{\Gamma_1}=J_\mu^5$ for $F_1$ and
$F_0$, and $j_{\Gamma_1}=J_\mu^{5T}$ for $F_T$.  The matrix element
of $j_{\Gamma_2}=\bar b i\gamma_5 s$ between the vacuum and $B_s$
defines the $B_s$ decay constant $f_{B_s}$ : $\langle \overline
B_s(p_{B_s})| \bar b i\gamma_5 s|0\rangle =
 \frac{m_{B_s}^2}{m_{b}+m_s}f_{B_s}$.

The LCSR method consists in evaluating the correlation function in
Eq.~(\ref{corr}) both at the hadron   level and at the quark level.
Equating the two representations allows us to obtain a set of sum
rules suitable to derive the form factors.

The hadronic representation of the correlation function in Eq.~
(\ref{corr}) can be written as the sum of the contribution of the $\bar B_s$ state
and that of the higher resonances and the continuum of states $h$:
\begin{eqnarray}
 \hspace{-1.4cm}&&\Pi^{\rm H}(p_{f_0},q)=  \frac{\langle {f_0}(p_{f_0})|j_{\Gamma_1}|\overline
 {B}_s( p_{f_0}+q)\rangle \langle \overline {B}_s(p_{f_0}+q)|j_{\Gamma_2}|0\rangle}
 {m_{B_s}^2-(p_{f_0}+q)^2}+ \int_{s_0}^\infty ds \frac{\rho^h(s,q^2)}
 {s-(p_{f_0}+q)^2}, \label{hadronic}
\end{eqnarray}
where higher resonances and  the continuum of states are described
in terms of the spectral function $\rho^h(s,q^2)$, contributing
above a threshold $s_0$.

The correlation function  can be evaluated in QCD with the
expression
\begin{eqnarray}
 \Pi^{\rm QCD}(p_{f_0},q)=   \frac{1}{\pi}\int_{(m_b+m_s)^2}^\infty ds \, \frac{{\rm Im}\Pi^{\rm QCD}(s,q^2)}
 {s-(p_{f_0}+q)^2} \,. \label{QCD-repr}
\end{eqnarray}
Expanding the T-product in  Eq.~(\ref{corr}) on the light-cone, we
obtain a series of operators, ordered by increasing twist, the
matrix elements of which between the vacuum and the $f_0$  are
written in terms of  $f_0$ light-cone distribution amplitudes (LCDA).
Since the hadronic spectral function $\rho^h$ in (\ref{hadronic}) is
unknown, we use the global quark-hadron duality to identify $\rho^h$
with $\rho^{\rm QCD}={1 \over \pi} {\rm Im} \Pi^{\rm QCD}$ when
integrated above $s_0$ so that $ \int_{s_0}^\infty  ds
{\rho^h(s,q^2) \over s-(p_{f_0}+q)^2}=\frac{1}{\pi}\int_{s_0}^\infty
ds \, \frac{{\rm Im}\Pi^{\rm QCD}(s,q^2)}{s-(p_{f_0}+q)^2}$. Using the
quark-hadron duality, together with  the equality $\Pi^{\rm
H}(p_{f_0},q)=\Pi^{\rm QCD}(p_{f_0},q)$ and performing a Borel
transformation of the two representations, we obtain a generic sum rule for the form factors
\begin{eqnarray}
   {\langle {f_0}(p_{f_0})|j_{\Gamma_1}|\overline
{B}_s(p_{B_s})\rangle \langle \overline
{B}_s(p_{B_s})|j_{\Gamma_2}|0\rangle}
 e^{-\frac{m_{B_s}^2}{M^2}}=
 \frac{1}{\pi}\int_{(m_b+m_s)^2}^{s_0}ds \,\,
 e^{-\frac{s}{M^2}} \, \, {\rm Im}\Pi^{\rm QCD}(s,q^2), \label{SR-generic}
\end{eqnarray}
where $Q^2=-q^2$, $p_{B_s}=p_{f_0}+q$ and $M^2$ is the Borel
parameter. The Borel transformation will
improve the convergence of the series in $\Pi^{\rm QCD}$  and for
suitable values of $M^2$ enhances the contribution of the low lying
states to $\Pi^{\rm H}$. Eq.~(\ref{SR-generic}) allows us to derive
the sum rules for $F_1$, $F_0$ and $F_T$, choosing  
$j_{\Gamma_1}=J_\mu^5$ or $j_{\Gamma_1}=J_\mu^{5T}$.

\begin{figure}[b]
\includegraphics[width=0.38\textwidth]{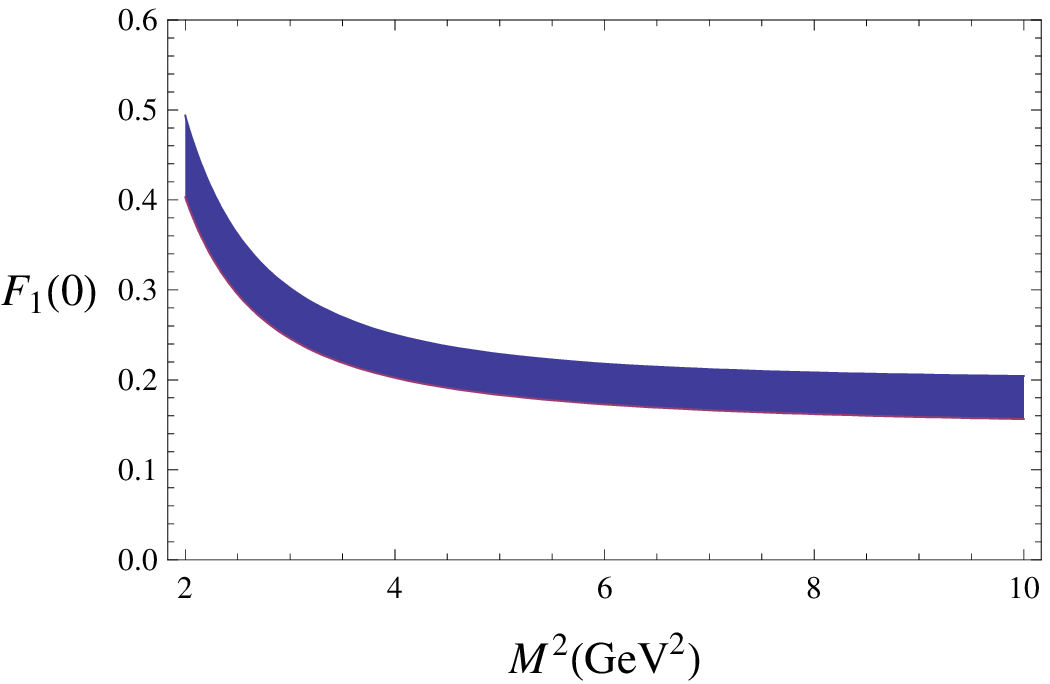}
\includegraphics[width=0.38\textwidth]{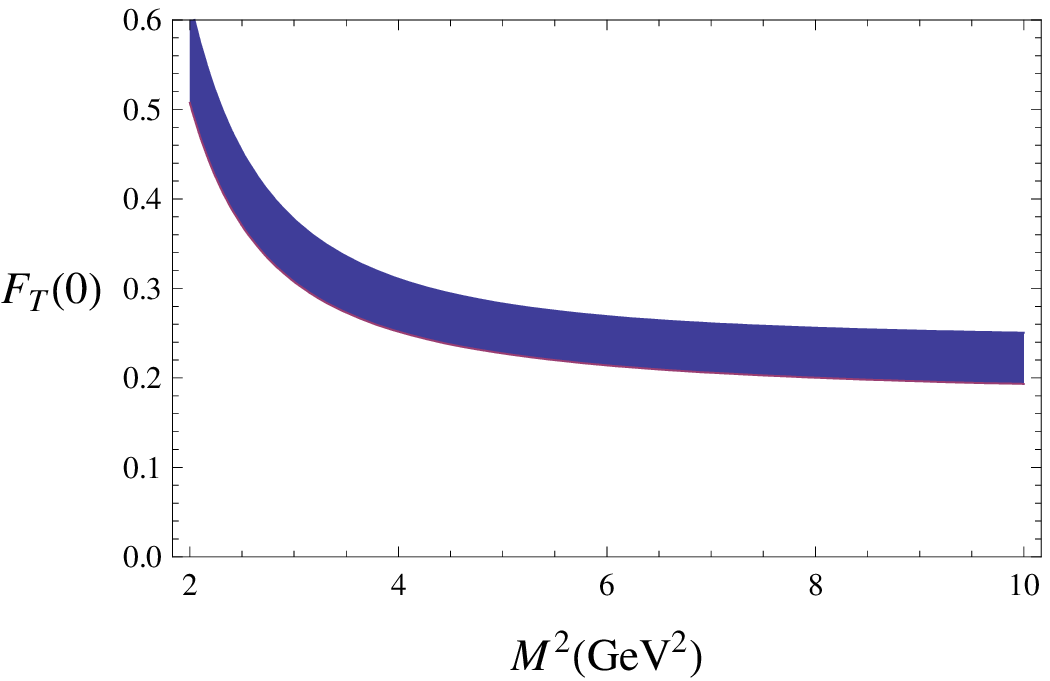}
\caption{Dependence of $F_1^{B_s \to f_0}(0)$ and $F_T^{B_s \to
f_0}(0)$ on the Borel parameter $M^2$. }\label{fig:M2dependence}
\end{figure}

We refer to Ref.~\cite{Colangelo:2010bg} for numerical values of the
input parameters as well as for the final expressions of the form
factors obtained from (\ref{SR-generic}). The $s_0$ is supposed to be
around the mass squared of the first radial excitation of $B_s$ and
is fixed as $s_0=(34\pm 2)\, {\rm GeV}^2$. As for the Borel
parameter, the result is obtained requiring stability against
variations of $M^2$. In Fig.~\ref{fig:M2dependence}  we show the
dependence of $F_1(q^2=0)$ and $F_T(q^2=0)$ on $M^2$ and we find the
stabilities when $M^2>6$ ${\rm GeV}^2$, and thus we choose
$M^2=(8\pm2)\, {\rm GeV}^2$.

To describe the form factors in the whole kinematically accessible
$q^2$ region, we use the parameterization
${F_i(q^2)=\frac{F_i(0)}{1-a_iq^2/m_{B_s}^2+b_i(q^2/m_{B_s}^2)^2}}$,
$ i \in \{1,0,T\}$; the parameters $F_i(0)$,  $a_i$ and $b_i$
are obtained through fitting the form factors computed numerically
in the large recoil region. Our results are collected in
Table~\ref{table:LO-formfactor}, where uncertainties in the results
are due to the input parameters, $s_0$ and $M^2$.
\begin{table}
\caption{$B_s \to f_0$  form factors in the LCSR.
}\label{table:LO-formfactor}
\begin{tabular}{ c c c c c }
\hline & $F_i(q^2=0)$  & $a_i$ & $b_i$
\\\hline
 $F_1 $   &   $0.185\pm0.029$
   &   $1.44^{+0.13}_{-0.09}$  &   $0.59^{+ 0.07}_{-0.05}$ \\
 $F_0 $   &   $0.185\pm0.029$
    &   $0.47^{+0.12}_{-0.09}$  &   $0.01^{+  0.08}_{-0.09}$\\
 $F_T $   &   $0.228\pm0.036$
    &   $1.42^{+0.13}_{-0.10}$  &   $0.60^{+  0.06}_{-0.05}$ \\
\hline
\end{tabular}
\end{table}

\section{$B_s \to M_{c{\bar c}} L$ decays}
\label{decays}

The effective hamiltonian responsible for decays induced
by the $b \to c {\bar c}s$ transition is:
 \begin{eqnarray}
 {\cal H}_{\rm eff} &=& \frac{G_{F}}{\sqrt{2}}
     \big\{  V_{cb} V_{cs}^{*} \big[
     C_{1}({\mu}) O_{1}
  +  C_{2}({\mu}) O_{2}\big] -V_{tb} V_{ts}^{*} \big[{\sum\limits_{i=3}^{10,7\gamma,8g}} C_{i}({\mu}) O_{i}({\mu})
  \big ] \big\} ,
 \label{eq:hamiltonian-b-ccs}
\end{eqnarray}
where $G_F$ is the Fermi constant, $O_i$ are the four-quark or magnetic-moment operators and
$C_i$ are Wilson coefficients. With the assumption of the CKM
unitarity and the neglect of the tiny $V_{ub}V_{us}^*$, we have
$V_{tb} V_{ts}^{*}=-V_{cb} V_{cs}^{*}$.

The simplest approach to
compute the matrix element of the effective four-quark or
magnetic-moment operators  is the naive factorization approach. 
In this approach neglecting the magnetic moment operators whose contributions are
suppressed by $\alpha_s$, the $B_a \to M_{c {\bar c}} L$
amplitude reads (($a$ being a light flavour index))
\begin{eqnarray}
{\cal A}(B_a \to M_{c {\bar c}} L)={G_F \over \sqrt{2}}
V_{cb}V_{cs}^* a_2^{\rm eff}(\mu) \langle{M_{ c{\bar c} }}| {\bar c}
\gamma^\mu(1-\gamma_5) c \, |{0}\rangle\langle{L} |{\bar s}
\gamma_\mu(1-\gamma_5) b |\bar {B}_a\rangle \,, \label{amplitude-compl}
\end{eqnarray}
where $a_2^{\rm eff}(\mu)$ is a combination of the Wilson
coefficients: $a_2^{\rm eff}(\mu)=a_2(\mu)+a_3(\mu)+a_5(\mu)$ and and
$a_2=C_2+{C_1 / N_c}$, $a_3=C_3+{C_4 / N_c}+{3 \over 2}e_c
\left(C_9+{C_{10}/ N_c} \right)$ and $a_5=C_5+{C_6 /N_c}+{3 \over
2}e_c \left(C_7+{C_{8} /N_c} \right)$ with $e_c=2/3$ and $N_c=3$. This factorization approach allows us to
express the decay amplitudes in terms of the heavy-to-light form factors and the decay constant of the emitted meson.
Unfortunately, one severe drawback is that naive factorization badly reproduces several
branching ratios for which experimental data are available. In
particular, the $b \to c{\bar c}s$ induced modes under scrutiny are
color suppressed, and the predictions of naive factorization will
typically undershoot the data. The most striking discrepancy is for
the $B_d$ decay modes with $\chi_{c0}$ in the final state, which have a sizeable
rate but their decay amplitude vanishes in the factorization
approach~\cite{Colangelo:2002mj}.

Several modifications to the  naive factorization ansatz have been
proposed and in particular in this work we will explore one
possibility by treating the Wilson coefficients as effective
parameters to be determined from experiment. In principle, it
implies that such coefficients are channel dependent. However for
channels related  by invoking flavour symmetries, universal values
for the coefficients can be assumed. In our case, this {\it
generalized} factorization approach consists in considering the
quantity $a_2^{\rm eff}$ in (\ref{amplitude-compl}) as a process
dependent parameter to be fixed from experiment. In particular, if
one assumes the flavor $SU(3)$ symmetry, $B$ ($B_u$ or $B_d$) decays can be
related to analogous $B_s$ decays, so that experimental data on $B$ decays provide a
prediction for $B_s$ related ones.

Our strategy is to exploit the existing experimental data for $B$
decay modes to determine an effective parameter $a_2^{\rm eff}$ and,
assuming $SU(3)_F$ symmetry, to use these values to predict the
flavour related $B_s$ decays. In the case of modes with $\chi_{c1}$ in the final state, we will determine the combination $f_{\chi_{c1}}a_2^{\rm eff}$ 
since sizable uncertainties may be introduced to the Wilson coefficient but will cancel in the predictions of branching ratios.  
In this procedure, we use two
sets of form factors: the one obtained using sum rules based on the
short-distance expansion
\cite{Colangelo:1995jv}, and the set in \cite{Ball:2004ye} based on
the light-cone expansion. In the case of $B_s \to \phi$  and $B_s
\to f_0(980)$ we use form factors determined by
LCSR~\cite{Ball:2004ye,Colangelo:2010bg}. $B_s\to
\eta^{(\prime)}$ form factors are related to the analogous $B \to K$
form factors and the mixing angle between $\eta$ and $\eta'$ in the
flavor basis~\cite{Feldmann:1998vh} can be fixed to the value
measured by the KLOE Collaboration: $\theta=\big( 41.5 \pm
0.3_{stat} \pm 0.7_{syst} \pm0.6_{th} \big )^\circ$  \cite{kloe},
which is also supported by a QCD sum rule analysis of the radiative
$\phi \to \eta^{(\prime)} \gamma$ modes \cite{DeFazio:2000my}. The
Wilson coefficients for $B_s \to M_{c {\bar c}} f_0(980)$  are
obtained using the effective value determined from $B \to M_{c {\bar
c}} K$. Most of other numerical inputs are taken from the particle
data group and we refer to Ref.~\cite{Colangelo:2010zz} for more
details.

The predictions for branching ratios of $B_s$ decays  are given in Tables
\ref{tab:Bs-BR} and \ref{tab:Bs-p-wave}. In Table \ref{tab:Bs-BR} the available experimental
data~\cite{Amsler:2008zzb,:2009usa,talkbelle} are also reported, with a
satisfactory agreement with the predictions. In theoretical predictions we have included the uncertainty on the form factors at
$q^2=0$ and on the experimental branching ratios, but in the case of
the modes involving $\eta$ or $\eta^\prime$ the uncertainty on the
form factors is not included since the dependence on the form factors will cancel when the branching ratios of $B\to J/\psi K$ decays are related to the corresponding $B_s$ decays. 
%
%

\begin{table}
\caption{Branching ratios (in units of $10^{-4}$) of $B_s\to
M_{c\bar c}\, L$ using the form factors in \cite{Colangelo:1995jv}
(CDSS) and in \cite{Ball:2004ye} (BZ). Experimental results are
taken from PDG \cite{Amsler:2008zzb}, except for $B_s \to J/\psi \,
\eta\,(\eta')$ measured by Belle Collaboration~\cite{:2009usa} and
the bound  for $B_s \to J/\psi \, f_0$ is from
Ref.~\cite{talkbelle}.\label{tab:Bs-BR}}
\begin{tabular}{cccccccc}
\hline mode                &      ${\cal B}$ (CDSS)      &
${\cal B}$ (BZ)                  & Exp. & mode                 &
${\cal B}$ (CDSS)              &${\cal B}$ (BZ)          \\\hline
 $J/\psi \, \eta$     & $4.3\pm 0.2$  & $4.2\pm 0.2$ &$3.32\pm1.02$  &  $\eta_c  \, \eta$  & $4.0\pm 0.7$  & $3.9\pm 0.6$ \\
 $J/\psi \, \eta'$    & $4.4\pm 0.2$  & $4.3\pm 0.2$ & $3.1\pm1.39$    &  $\eta_c  \, \eta'$ & $4.6\pm 0.8$  & $4.5\pm 0.7$ \\
  \\
 $\psi(2S)\, \eta$ & $2.9\pm 0.2$  &$3.0\pm 0.2$  &                             &$\eta_c(2S) \, \eta$  &$1.5\pm 0.8$     & $1.4\pm 0.7$ \\
 $\psi(2S) \, \eta'$& $2.4\pm 0.2$  & $2.5\pm 0.2$ &                             &$\eta_c(2S) \, \eta'$ &$1.6\pm 0.9$     & $1.5\pm 0.8$ \\
 \\
 $J/\psi\, \phi$      & ---                      &$16.7\pm 5.7$&$13\pm4$          &$\eta_c \, \phi$ &--- &  $15.0\pm 7.8$ \\
 $\psi(2S)\, \phi$ &---                       &$8.3\pm 2.7$   &$6.8\pm3.0$      &                       &     &\\
 \\
$\chi_{c1} \, \eta$ &$2.0\pm 0.2$ &$2.0\pm 0.2$ & &  $\chi_{c1}\, f_0$ &     $1.88\pm0.77$    & $0.73\pm0.30$ \\
 $\chi_{c1} \, \eta'$ &$1.9\pm 0.2$ &$1.8\pm0.2$ & &  $\chi_{c1} \, \phi$ & ---  & $3.3\pm1.3$\\
 \hline
$J/\psi \, f_0$    & $4.7\pm 1.9$   & $2.0\pm 0.8$ & $<3.26$ &  $\eta_c \, f_0$      & $4.1\pm 1.7$  & $2.0\pm 0.9$ \\
$\psi(2S) \, f_0$  & $2.3\pm 0.9$  &$0.89\pm0.36$ &    & $\eta_c(2S)
\, f_0$  & $0.58\pm 0.38$& $1.3\pm 0.8$  \\ \hline \end{tabular}
\end{table}

As appears from
Tables~\ref{tab:Bs-BR} and \ref{tab:Bs-p-wave} all the considered
modes have sizable branching fractions which are large enough to make them promising
candidates for the measurement of  $\beta_s$. The modes involving
$\eta, \,\eta^\prime, \, f_0$ present, with respect to the golden
mode $B_s\to J/\psi\phi$, the advantage that the final state is a CP
eigenstate, not requiring angular analysis. However,  
channels with $\eta$ and $\eta'$ can be  useful only after a
number of events have been accumulated, since at least two photons are
required for the reconstruction.
\begin{table}
\caption{Branching ratios of $B_s$ decays
into p-wave charmonia (unit: $10^{-4}$). \label{tab:Bs-p-wave}}
\begin{tabular}{ccccccccccc}   \hline mode         & ${\cal B}$      & mode &
${\cal B}$     & mode          & ${\cal B}$\\ \hline
 $\chi_{c0}\, \eta$ & $0.85\pm0.13$ & $\chi_{c2}\, \eta$   & $<0.17$        & $h_c\, \eta$   & $<0.23$\\
 $\chi_{c0}\, \eta'$& $0.87\pm0.13$  & $\chi_{c2}\, \eta'$  & $<0.17$        & $h_c\, \eta'$  & $<0.23$\\
 $\chi_{c0}\, f_0$& $1.15\pm0.17$   & $\chi_{c2}\, f_0$  & $<0.29$        & $h_c\, f_0$  & $<0.30$\\
 $\chi_{c0}\, \phi$ & $1.59\pm0.38$ & $\chi_{c2}\, \phi$   & $<0.10(0.62\pm0.17)$ & $h_c\, \phi$   & $(<1.9)$ \\ \hline
\end{tabular}
\end{table}

As discussed  in \cite{Stone:2008ak,Colangelo:2010bg,Leitner:2010fq}, $B_s\to
J/\psi f_0$ has  appealing features since, compared with the
$\eta^{(')}$, the $f_0$ can be easily identified in the $\pi^+\pi^-$
final state with a large BR: ${\cal B}(f_0\to
\pi^+\pi^-)=(50^{+7}_{-8})\%$~\cite{Ablikim:2004cg}, so that  this channel can likely be
accessed. At present, the Belle Collaboration has recently provided
the following upper limit \cite{talkbelle}:
\begin{eqnarray}
{\cal B}(B_s\to J/\psi f_0)\times {\cal B}(f_0\to \pi^+\pi^-)<
1.63\times 10^{-4}\,
\end{eqnarray}
marginally in accordance  with our prediction.
 
Let us come to $B_s$ decays to  $p$-wave charmonia. Among these
decays,  the only one with non vanishing amplitude in the
factorization assumption is that  with $\chi_{c1}$ in the final
state. In the other cases, i.e. modes involving $\chi_{c0,2}$ or
$h_c$, which we  show in Table \ref{tab:Bs-p-wave},  results are obtained
determining the decay amplitudes from the $B$ decay data by making
use of the SU(3) symmetry. In this case, the differences between the
$B$ and $B_s$ decays arise from the phase space and lifetimes of the
heavy mesons. As for the mechanism inducing such processes, one
possibility is that
rescattering may be responsible of their observed branching
fractions, as proposed  in Ref.\cite{Colangelo:2002mj}. Among these channels,  $B_s\to \chi_{c0}\phi$ is of prime
interest and promising for both hadron colliders and $B$
factories.

%

\section{conclusion}
Recent results in the $B_s$ sector strongly require theoretical
efforts to shed light on which are the most promising decay modes to
unreveal new physics. In this work we have analyzed channels
induced by the $b \to c {\bar c}s$ transition. Modes with a
charmonium state plus $\eta$, $\eta^\prime$, $f_0(980)$ are the most
promising, being CP eigenstates not requiring an angular analysis. In
particular, the case of $f_0$ is particularly suitable in view of
its easier reconstruction in the subsequent decay to $\pi^+ \pi^-$.
As a preliminary step we have used the light-cone sum rules to compute the $B_s\to f_0(980)$ form
factors which are necessary inputs in the analysis of $B_s$ decays.


\begin{theacknowledgments}
I thank the workshop organizers for the nice week in Martina Franca.
I also thank P. Colangelo and F. De Fazio for collaboration.
\end{theacknowledgments}



\bibliographystyle{aipproc}   

%


\end{document}